\documentclass{article}
\usepackage{ijcai26}
\usepackage{times}
\usepackage{url}
\usepackage{named}
\usepackage{amsmath,amssymb,amsfonts}
\usepackage{algorithmic}
\usepackage{algorithm}
\usepackage{graphicx}
\usepackage{textcomp}
\usepackage{xcolor}
\usepackage{booktabs}
\usepackage{enumitem}

\setlist[itemize,enumerate]{nosep,leftmargin=*}

\title{ContestTrade: A Multi-Agent Trading System Based on Internal Contest Mechanism}

\author{
    Li Zhao$^1$, Rui Sun$^1$, Zuoyou Jiang$^1$, Bo Yang$^1$, \\
    Yuxiao Bai$^2$, Mengting Chen$^2$, Jing Li$^1$, Zuo Bai$^{1,2}$
    \affiliations
    $^1$Stepfun \\
    $^2$FinStep
    \emails
    baizuo@stepfun.com; baizuo@finstep.cn
}

\begin{document}

\maketitle

\begin{abstract}
In financial trading, large language model (LLM)-based agents demonstrate significant potential, but their decisions can be sensitive to noisy and non-stationary market information. We propose \textbf{ContestTrade}, a multi-agent trading system with an internal competitive mechanism inspired by institutional investment workflows. The system consists of two specialized teams:
(1) a \textbf{Data Team} that processes and condenses massive market data into diversified textual factors optimized for constrained LLM context windows, and
(2) a \textbf{Research Team} that produces parallelized multipath trading decisions via tool-augmented deep research.
The core design is a ``Quantify-Predict-Allocate'' contest mechanism within each team: agent outputs are scored only after market outcomes become observable, future utility is predicted from historical scores, and resources are allocated to agents with positive predicted utility.
In a post-2024 A-share backtest, ContestTrade achieves higher backtested return and risk-adjusted performance than the evaluated baselines. We further describe the temporal protocol, implementation choices, and limitations to clarify the scope of these results.
\end{abstract}

\section{Introduction}
\label{sec:introduction}

The financial sector is undergoing a profound transformation with the rise of LLM-based agents \cite{wu2023bloomberggpt,huang2023finbert}. These agents excel at processing complex market information, offering interpretable outputs through natural language explanations and, when integrated with external tools, can flexibly incorporate diverse information sources including news, numerical data, and sentiment indicators. Recent advancements highlight LLMs' potential to automate complex trading decisions and achieve competitive performance in dynamic markets \cite{lopez2024gptforcast,fatouros2024beatwallstreet,zhang2024unveiling}.

However, market volatility and noise \cite{malkiel1973random,engle1982arch} pose significant challenges. High sensitivity to noise often leads to inconsistent decision-making and undermines performance. Traditional single-agent approaches struggle to capture intricate temporal dependencies and resolve conflicting signals, especially during market turbulence.

Awareness of these challenges has fueled interest in multi-agent systems \cite{lebaron2006multiagent}, leveraging role specialization for enhanced robustness \cite{byrd2019abides}. Despite advances, current multi-agent frameworks face key limitations: fixed data pipelines that struggle to adapt to shifting market regimes, decisions based solely on individual agents' historical returns, and insufficient quantitative reasoning tools for complex market scenarios.

To address these limitations, we propose \textbf{ContestTrade}, a multi-agent trading system based on an internal contest mechanism. Unlike debate-style systems that aggregate agent opinions statically, ContestTrade treats each data analyst and research agent as an expert whose usefulness is estimated from delayed market feedback under a fixed trading protocol. The framework combines competitive evaluation with a Deep Research methodology that allows agents to use temporally constrained financial tools. Our main contributions are:
\begin{enumerate}
    \item We adapt tool-augmented Deep Research to financial trading, enabling LLM agents to plan multi-step investigations over news, fundamentals, technical indicators, and market data while respecting data-availability time stamps.
    \item We introduce a two-level Quantify-Predict-Allocate contest that uses realized historical feedback to select textual factors and weight research agents, connecting multi-agent trading to online expert selection and portfolio allocation.
    \item We provide a reproducible daily backtesting protocol for an A-share setting, including trading constraints, anti-leakage rules, and ablations that isolate the effect of data-level and researcher-level contests.
\end{enumerate}

\begin{figure*}[t]
    \centering
    \includegraphics[width=0.85\textwidth]{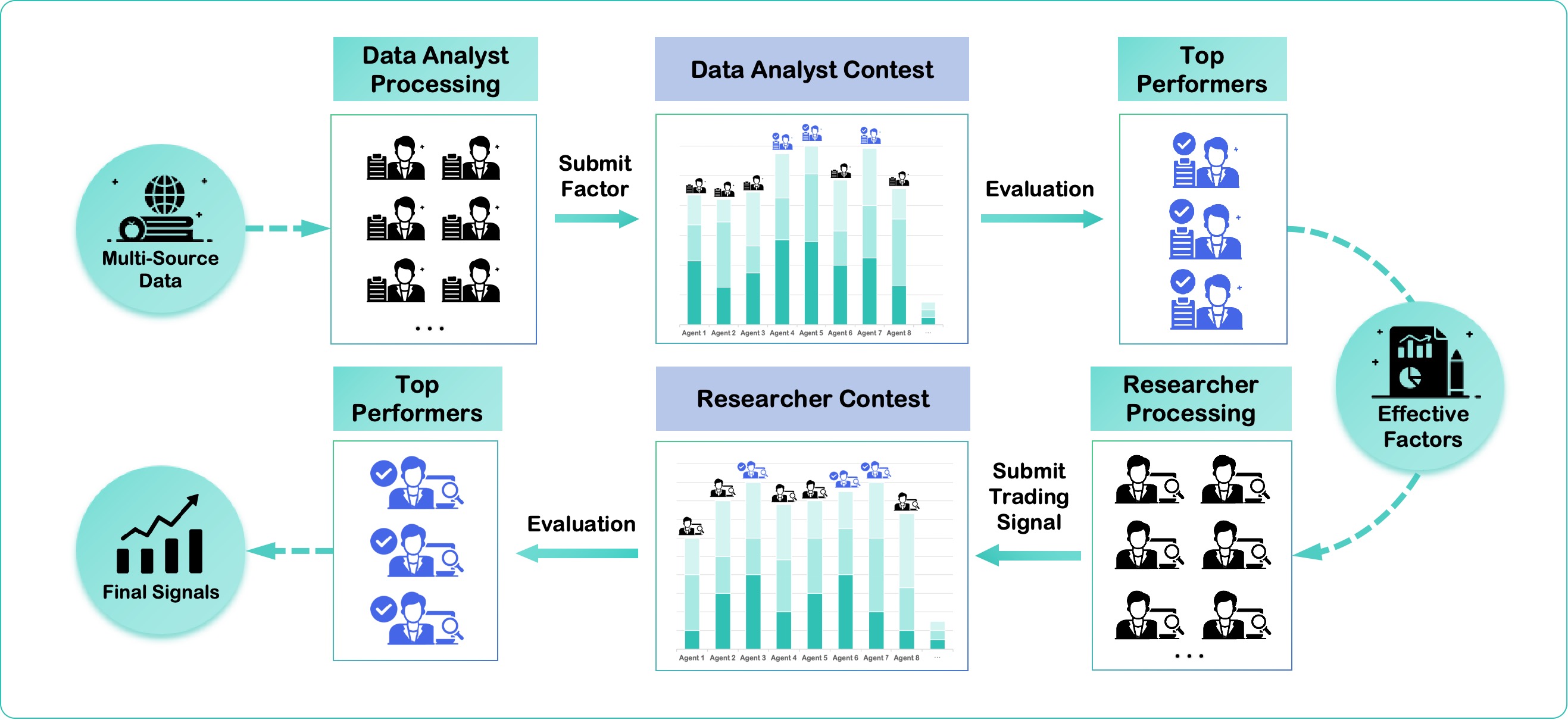}
    \caption{The ContestTrade Framework Architecture, showing the complete pipeline from multi-source data input through the Data Team and Research Team to final trading signals. Each team implements an internal contest mechanism that continuously evaluates and ranks agents based on market feedback.}
    \label{fig:framework_architecture}
\end{figure*}

\section{Related Works}
\label{sec:related_works}

\paragraph{LLM-Based Trading Agents.}
LLMs are revolutionizing trading workflows as autonomous decision-making agents. LLMFactor \cite{wang2024llmfactor} extracts factors from text for explainable stock movement prediction. FinGPT \cite{yang2023fingpt} and FinRobot \cite{yang2024finrobot} provide open-source resources, while TradingGPT \cite{li2023tradinggpt} emulates human cognition for trading. The SEP framework \cite{koa2024sep} enables explainable predictions via self-reflection. As alpha miners, QuantAgent \cite{wang2024quantagent} refines knowledge through real-world testing, while AlphaGPT \cite{wang2023alphagpt} pioneers ``Human-in-the-Loop'' strategies for effective alpha generation.

\paragraph{Adaptive and Multi-Agent Systems.}
A critical challenge is agent-level adaptability in volatile markets. FinMem \cite{yu2023finmem} introduces a memory module for continuous decision refinement. FinAgent \cite{zhang2024finagent} uses dual-level reflection for rapid adaptation. Beyond individual capabilities, HAD \cite{xing2024had} and TradingAgents \cite{xiao2024tradingagents} employ specialized agents that collaborate through discussions and debates. FinCon \cite{yu2024fincon} implements a manager-analyst hierarchy with dual-level risk control. The MASS framework \cite{guo2025mass} leverages large-scale multi-agent simulation for market understanding through emergent behaviors.

\paragraph{Tool-Augmented LLM Agents.}
Recent work has shown that augmenting LLMs with external tools significantly enhances their capabilities in complex reasoning tasks. The ReAct framework \cite{yao2023react} interleaves reasoning and action, enabling agents to dynamically retrieve and process external information. Toolformer \cite{schick2023toolformer} demonstrates that LLMs can learn to invoke APIs autonomously. In finance, tool augmentation is particularly critical: agents must access real-time market data, execute quantitative computations, and retrieve domain-specific knowledge that lies beyond static training corpora. Our Deep Research mechanism builds on this paradigm by equipping Research Agents with a comprehensive financial toolkit---including fundamental analysis APIs, technical indicator calculators, and news retrieval systems---enabling them to conduct multi-step investigative analysis rather than relying solely on parametric knowledge. This tool-augmented design is essential for bridging the gap between LLM reasoning capabilities and the quantitative rigor demanded by financial decision-making.

\section{Architecture}
\label{sec:architecture}

As illustrated in Figure~\ref{fig:framework_architecture}, ContestTrade operates through a structured dual-stage pipeline emulating investment firm dynamics. The architecture comprises two specialized teams: the Data Team and the Research Team.

The framework begins with Data Agents processing raw market data into textual factors. Our internal contest mechanism evaluates each agent after the corresponding market outcome becomes observable, forecasts near-term utility from past scores, and allocates context or capital accordingly. The system constructs an optimized factor portfolio by aggregating outputs with positive predicted efficacy, which is then passed to Research Agents for parallel analysis and a second stage of competition. The final portfolio is a Sharpe-weighted ensemble of qualified Research Agent proposals rather than an unconstrained free-form recommendation. This dual-stage competitive framework is designed to reduce exposure to persistently noisy agents while preserving diversity across independently generated views.

\subsection{Data Team Design}
\label{subsec:data_team}

\begin{figure}[t]
    \centering
    \includegraphics[width=0.48\textwidth]{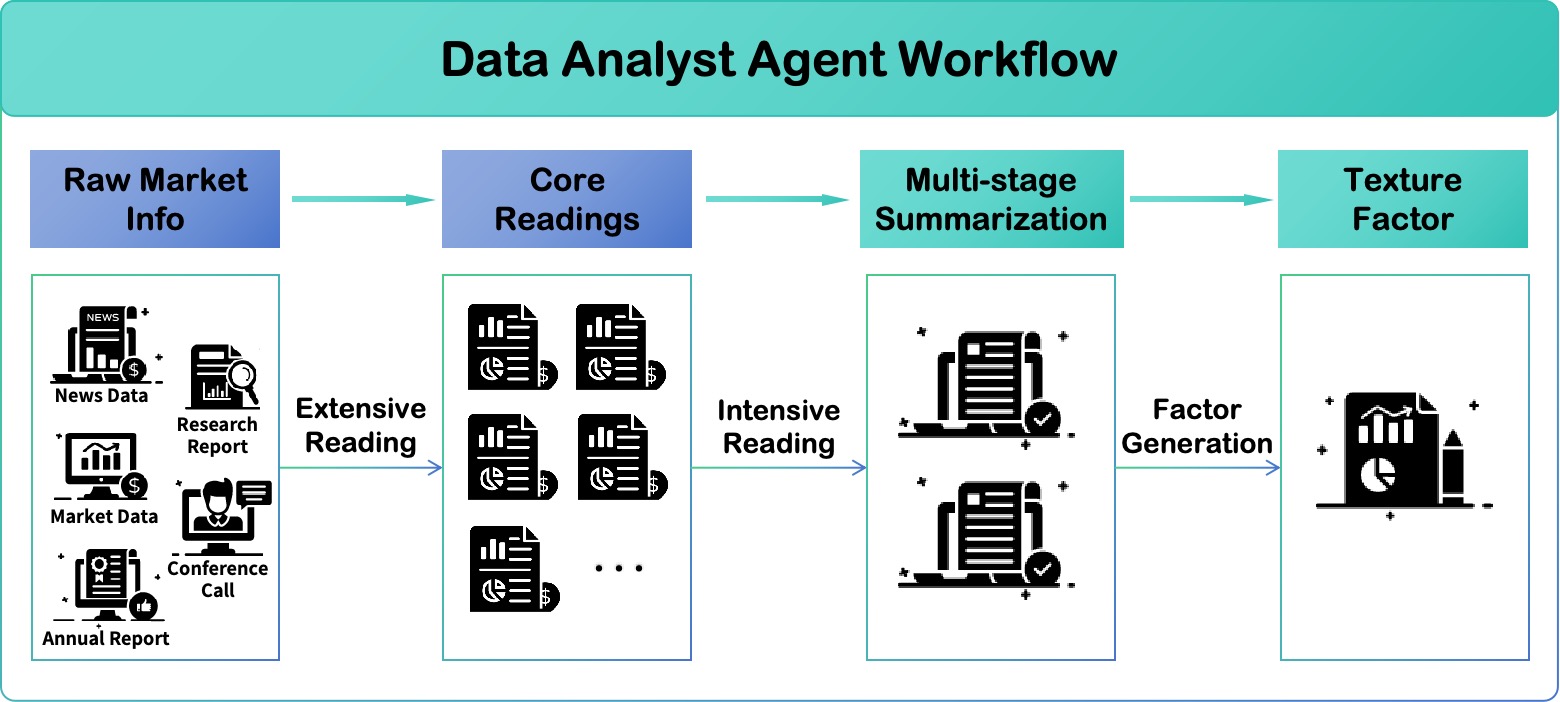}
    \caption{The Data Team Architecture, showing the workflow from raw market data to textual factor generation.}
    \label{fig:data_team}
\end{figure}

The Data Team distills vast volumes of raw market data into high-quality, context-friendly textual factors optimized for the constrained context windows of LLMs.

\subsubsection{Team Composition and Workflow}
The Data Team comprises multiple Data Analysis Agents operating in parallel (Figure~\ref{fig:data_team}). Each agent follows a three-stage workflow: (1) \textit{Dynamic Information Prioritization}---generating a daily focus to filter relevant news, financials, and announcements; (2) \textit{Parallel Intensive Reading}---leveraging LLM capabilities for end-to-end information synthesis; and (3) \textit{Textual Factor Generation}---producing a context-engineered summary capped at 4k tokens. The collective output forms a set of independent textual factors, evaluated through the internal contest mechanism to construct an optimized portfolio for the Research Team.

\subsection{Research Team Design}
\label{subsec:research_team}

\begin{figure}[t]
    \centering
    \includegraphics[width=0.48\textwidth]{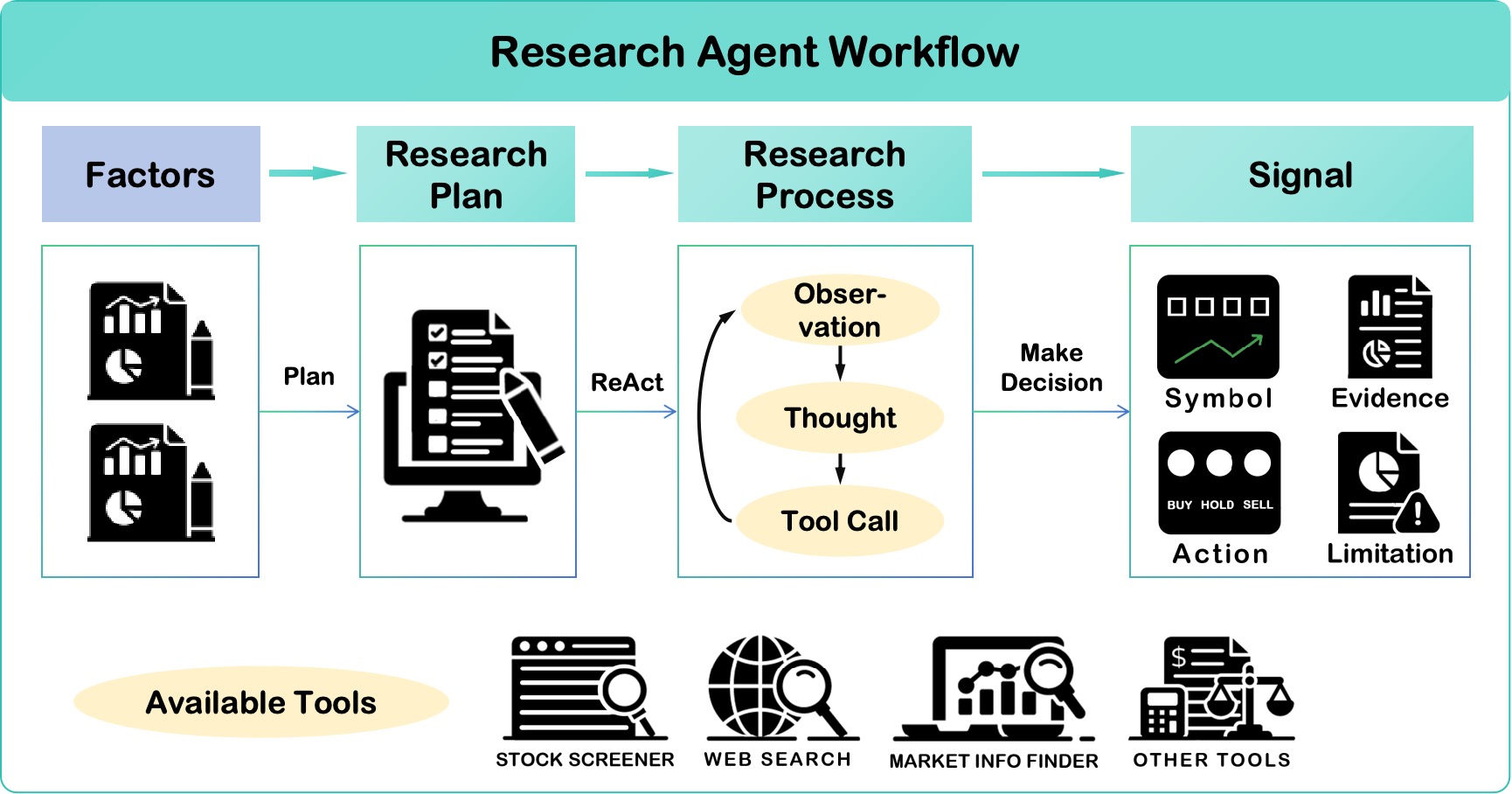}
    \caption{The Research Team Architecture, showing the workflow from textual factors to trading signal generation.}
    \label{fig:trading_strategy_team}
\end{figure}

The Research Team generates precise trading recommendations by leveraging the textual factors from the Data Team and conducting further in-depth analysis through parallelized multipath trading decisions.

\subsubsection{Team Composition and Workflow}
The Research Team comprises multiple autonomous Research Agents, each initialized with distinct ``trading beliefs'' generated by LLMs conditioned on predefined trading principles such as momentum, reversal, fundamentals, event-driven catalysts, and risk control. As illustrated in Figure~\ref{fig:trading_strategy_team}, each agent follows a Plan + ReAct \cite{yao2023react} framework: (1) \textit{Initial Planning}---receiving textual factors and planning trading opportunities; (2) \textit{Information Gathering}---leveraging specialized financial tools for deeper market analysis; and (3) \textit{Signal Generation}---producing structured trading signals with symbol, action, confidence, evidence, risk factors, and expected holding horizon. Agents are equipped with eight tools including web search, news retrieval, market data, and stock analysis. Every tool call is parameterized by the decision date, and the returned records are filtered by release time so that an agent cannot access information published after portfolio formation.

\subsection{Contest Mechanism}
\label{subsec:contest_mechanism}

\subsubsection{The Quantify-Predict-Allocate Framework}
The core of ContestTrade is an internal contest mechanism designed to enhance system adaptivity. Its objective is to channel resources towards agents with proven effectiveness. This mechanism is formalized as a three-phase ``Quantify-Predict-Allocate'' model:
\begin{equation} \label{eq:pipeline}
\mathcal{A} \xrightarrow{f_{\text{quant}}} \{ q_{i,t} \} \xrightarrow{f_{\text{predict}}} \{ \hat{u}_{i,t+n} \} \xrightarrow{\pi_{\text{allocate}}} \mathcal{W}_t
\end{equation}
This pipeline first quantifies each agent's historical performance yielding scores $\{q_{i,t}\}$ at time $t$, then predicts the future utility $\{\hat{u}_{i,t+n}\}$ over the next $n$ steps, and finally allocates resources based on these predictions. This mechanism provides a unified foundation for both teams:
\begin{enumerate}
    \item \textbf{Data Analyst Contest:} Agents compete to construct an optimal information portfolio for the Research Agents.
    \item \textbf{Researcher Contest:} Agents compete to achieve optimal capital allocation for final trading.
\end{enumerate}

\subsubsection{Data Analyst Contest}
\paragraph{Optimization Objective.}
The objective is to construct an optimal factor portfolio $\mathcal{F}_t$ from all available factors $\mathbb{F}_t$ to maximize the Research Agent's Decision Value ($DV$) \cite{chroma2024contextrot}, modeled as a product of Information Value ($V$) and Decision Capability ($DC$):
\begin{equation} \label{eq:dv_definition}
DV(\mathcal{F}_t)=V(\mathcal{F}_t) \cdot DC \quad \text{where} \quad V(\mathcal{F}_t)=\sum_{i \in \mathcal{F}_t}v_i
\end{equation}
Long-context studies suggest that an LLM's effective use of evidence can decay as context length $L$ grows \cite{liu2024lost,modarressi2025nolima}. We therefore use a simple sigmoid-shaped context penalty as an empirical design heuristic rather than a model-specific theorem:
\begin{equation} \label{eq:dc_sigmoid}
DC(L)=\frac{1}{1+e^{k(L-L_0)}}
\end{equation}
The optimization objective becomes selecting $\mathcal{F}_t \subseteq \mathbb{F}_t$ that maximizes:
\begin{equation} \label{eq:full_optimization}
\max_{\mathcal{F}_t \subseteq \mathbb{F}_t} \left( \sum_{i \in \mathcal{F}_t} v_i \right) \cdot \frac{1}{1+e^{k\left(\sum_{i \in \mathcal{F}_t} l_i - L_0\right)}}
\end{equation}
where $v_i$ and $l_i$ denote the latent value and length of factor $i$. In practice, we choose an effective context budget $L^* < L_0$ to leave room for instructions and Research Agent reasoning. The remaining problem is: (1) finding a quantifiable proxy $q_i$ for the latent value $v_i$, and (2) developing an allocation policy $\pi_t$ to form a compact, non-redundant factor portfolio.

\paragraph{Quantification via Zero-Intelligence Trader.}
To quantify a factor's latent value with an objective proxy $q_{i,t}$, we simulate a Zero-Intelligence (ZI) Trader \cite{gode1993allocative}. The score is computed only after the next-day return is realized and is used for future contest updates, not for the same day's trading decision. This delayed-update rule separates evaluation from decision-time information and avoids directly injecting future returns into current signals. The trader operates on each atomic ``Observation'' extracted from the factor under limited analysis and no external context:
\begin{equation} \label{eq:zi_score}
q_{i,t} = \sum_{\text{obs} \in F_{i,t}} \text{ZI}(\text{obs})
\end{equation}
where $F_{i,t}$ is the set of observations comprising factor $i$ at time $t$, and $\text{ZI}(\cdot)$ is detailed in Algorithm~\ref{alg:zi_trader}.

\begin{algorithm}[t]
\caption{ZI Trader}
\label{alg:zi_trader}
\begin{algorithmic}[1]
\REQUIRE A single Observation \textit{obs}
\ENSURE The quantified value (reward) for \textit{obs}
\STATE \textit{obs\_reward} $\leftarrow 0$
\STATE \textit{RatedSymbols} $\leftarrow$ \texttt{GroundAndRate}(\textit{obs})
\STATE \textit{// Each rating is an integer in} $\{-2, -1, 0, 1, 2\}$
\FOR{each $s$ in \textit{RatedSymbols}}
    \STATE \textit{reward} $\leftarrow s$.\textit{rating} $\times$ \texttt{PriceChange}($s$.\textit{code}, $t+1$)
    \STATE \textit{obs\_reward} $\leftarrow$ \textit{obs\_reward} + \textit{reward}
\ENDFOR
\RETURN \textit{obs\_reward}
\end{algorithmic}
\end{algorithm}

\paragraph{Prediction.}
Financial markets exhibit style rotation, rendering static policies suboptimal. We hypothesize that factor performance exhibits short-term momentum and evaluate this hypothesis with the Rank Information Coefficient (RIC):
\begin{equation}\label{eq:ric_comparison}
\text{RIC}(\overline{q}_{t-m:t}, \overline{q}_{t:t+n}) \gg \text{RIC}(\overline{q}_{t-M:t}, \overline{q}_{t:t+N})
\end{equation}
where $M \gg m, N \gg n$. The correlation is stronger for short-term windows in our validation period (we use $m{=}5, n{=}3$) and decays over longer horizons. We train a LightGBM model in a rolling manner: on each decision day $t$, the training set contains only samples whose labels have been realized strictly before $t$. The model maps $x_{i,t}=\Phi(q_{i,t-m+1:t})$---including trailing mean, volatility, recent trend, and drawdown of each factor score---to predicted future score $\hat{\mu}_{i,t+n}$ and volatility $\hat{\sigma}_{i,t+n}$, defining predicted utility as $\hat{u}_{i,t+n} = \hat{\mu}_{i,t+n} / \hat{\sigma}_{i,t+n}$.

\paragraph{Allocation.}
Textual factors often exhibit information redundancy. We model the portfolio's aggregate value as a submodular set function \cite{krause2014submodular}. For $f: 2^{\mathbb{F}_t} \rightarrow \mathbb{R}$ with submodularity guarantees, the optimization becomes:
\begin{equation} \label{eq:submodular_allocation}
\mathcal{F}_t^* = \underset{\mathcal{F}_t \subseteq \mathbb{F}_t}{\text{argmax}} \ f(\mathcal{F}_t) \quad \text{s.t.} \quad \sum_{i \in \mathcal{F}_t} l_i \le L^*
\end{equation}
We instantiate $f$ using a facility-location objective with non-negative utilities:
\begin{equation} \label{eq:facility_location}
f(\mathcal{F}_t) = \sum_{j \in \mathbb{F}_t} \max_{i \in \mathcal{F}_t} \max(0,\hat{u}_{i,t+n}) \cdot \text{sim}(i, j),
\end{equation}
where $\text{sim}(i,j)$ is the cosine similarity between text embeddings of two factors. Negative-utility factors are excluded from the coverage gain, which preserves monotonicity for the greedy approximation. We set $L_0{=}32\text{k}$ and $L^*{=}16\text{k}$ as engineering budgets for our model and prompt template, and solve the knapsack-constrained objective with a lazy greedy heuristic \cite{minoux1978accelerated}. The portfolio is reconstructed every $n$ days.

\subsubsection{Researcher Contest}
\paragraph{Optimization Objective.}
The objective is to dynamically allocate capital among research agents to maximize the portfolio's future risk-adjusted return.

\paragraph{Quantification via Hybrid Assessment.}
We construct a judger-augmented performance score $q_{i,t}$ composed of: (1) \textbf{Realized Performance}---standard quantitative metrics over a trailing $m$-day window (e.g., realized Sharpe Ratio), and (2) \textbf{Judgmental Quality}---qualitative scores from an LLM Judger Panel assessing logical soundness, evidence quality, risk awareness, and consistency between recommended actions and cited evidence. The judging prompt hides agent identifiers and uses a fixed rubric from 1 to 5 for each dimension; the final judgmental score is the average across panel members. This qualitative term is used only as a complement to realized performance and cannot override negative realized utility by itself.

\paragraph{Prediction.}
Agent performance exhibits short-term momentum in our validation data, and we use a prediction window of $n{=}5$ days for strategies (longer than $n{=}3$ for factors), aligning with the intuition that reasoned strategies possess greater performance inertia. As in the Data Analyst Contest, LightGBM is trained with a rolling historical split and maps $q_{i,t-m+1:t}$ to predicted Sharpe Ratio $\hat{u}_{i,t+n}$.

\paragraph{Allocation.}
We employ Predicted Sharpe Ratio-Weighted allocation, assigning capital proportionally to agents with positive predicted Sharpe Ratios. The ensemble first converts each selected agent's structured signal into target stock weights; conflicting long/flat recommendations are netted, and final capital weights are normalized under the trading constraints described in Section~\ref{subsec:experiment_setup}:
\begin{equation} \label{eq:weight_allocation}
w_{i,t} = \frac{\max(0, \hat{u}_{i,t+n})}{\sum_{j=1}^{N} \max(0, \hat{u}_{j,t+n})}
\end{equation}

\section{Experiments}
\label{sec:experiments}

\subsection{Experiment Setup}
\label{subsec:experiment_setup}

Our experiments utilize a real-world financial dataset encompassing news, corporate financials, announcements, and market data. The testing period is January--June 2025, and all internal model training and validation use data from July--December 2024 or earlier realized contest labels. Because LLM knowledge cutoffs alone cannot rule out backtest leakage, we enforce a decision-time data protocol: for a portfolio formed after market close on day $t$, the system can access only records whose public release timestamp is no later than the formation time; realized returns from $t{+}1$ or later are used only to update future contest scores after they become observable. Trading simulations are conducted at daily frequency on the A-share market, adhering to T+1 settlement, daily price limits, and 0.001 transaction cost.

The tradable universe is constructed from liquid A-share stocks available during the backtest window after excluding ST stocks, newly listed stocks with insufficient history, suspended stocks on the trading day, and stocks that cannot be traded because of price-limit constraints. Portfolio formation uses close-to-close daily returns, maintains cash for unfilled orders, and applies the same transaction-cost and execution assumptions to all methods. These rules are intended to make the comparison conservative, although we do not claim that the backtest captures full market impact or capacity constraints.

We benchmark against diverse strategies: Broad Market Index (CSI ALL Share), Rule-based Methods (MACD, RSI\&KDJ), Machine Learning (LGBM), Deep Learning (LSTM), Deep Reinforcement Learning (A2C, PPO), and Multi-Agent Systems (MASS \cite{guo2025mass}). All baselines use the same tradable universe, test period, transaction cost, and trading constraints. The standalone LGBM baseline is a price-feature cross-sectional predictor trained without textual factors or contest scores, whereas ContestTrade's LightGBM modules predict agent utility from historical contest scores; the two models therefore answer different subproblems and should not be interpreted as identical LightGBM configurations.

For LLM configuration, we use DeepSeek-V3 \cite{liu2024deepseek} as the backbone LLM for reproducibility. Data Analysis Agents and Research Agents use DeepSeek-V3, with Research Agents switching to DeepSeek-R1 for critical signal generation due to its enhanced reasoning capabilities. Prompts are fixed across the test period; sampling temperature is kept low for extraction and judging, and higher only for generating diversified trading beliefs. Failed tool calls are retried once and otherwise treated as missing evidence rather than replaced with later information.

We employ two categories of metrics: (1) \textbf{Strategy Performance Metrics}---Cumulative Return (CR), Sharpe Ratio (SR), and Maximum Drawdown (MDD); and (2) \textbf{Contest Effectiveness Metrics}---Rank Information Coefficient (Rank IC) and ICIR, which evaluate the predictive quality of our internal contest mechanisms. Given the six-month evaluation horizon, we present these results as an initial controlled backtest rather than a full claim of market-cycle generalization.

\subsection{Main Results}
\label{subsec:main_results}

\begin{table}[t]
\centering
\renewcommand{\arraystretch}{1.2}
\caption{Strategy performance comparison. Best results in \textbf{bold}.}
\label{tab:strategy_comparison}
\begin{tabular}{l ccc}
\toprule
\textbf{Model} & \textbf{CR (\%)} & \textbf{SR} & \textbf{MDD (\%)} \\
\midrule
CSI ALL Share & 4.42 & 0.46 & 13.75 \\
\midrule
\multicolumn{4}{l}{\textit{Rule-based Methods}} \\
MACD & 2.69 & 0.10 & 10.65 \\
RSI\&KDJ & 8.19 & 0.47 & 8.30 \\
\midrule
\multicolumn{4}{l}{\textit{ML/DL-based Methods}} \\
LGBM & -25.94 & -1.30 & 34.17 \\
LSTM & 8.34 & 0.51 & 29.56 \\
\midrule
\multicolumn{4}{l}{\textit{DRL-based Methods}} \\
A2C & 7.89 & 0.69 & 18.84 \\
PPO & 15.07 & 1.33 & 17.11 \\
\midrule
\multicolumn{4}{l}{\textit{Multi-Agent Methods}} \\
MASS & -19.12 & -1.76 & 24.55 \\
\midrule
\textbf{ContestTrade (Ours)} & \textbf{52.80} & \textbf{3.12} & \textbf{12.41} \\
\bottomrule
\end{tabular}
\end{table}

\begin{figure}[t]
    \centering
    \includegraphics[width=0.48\textwidth]{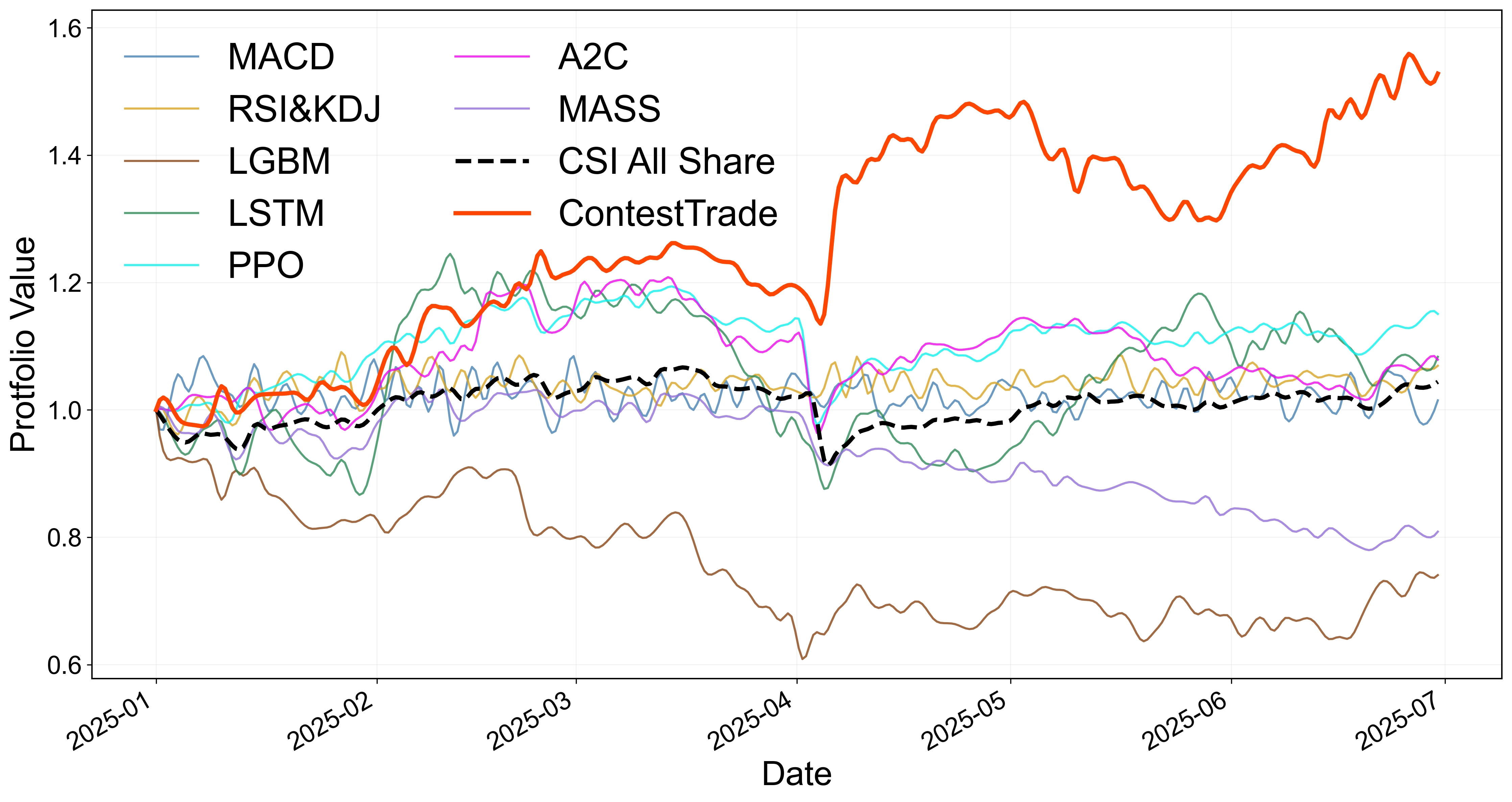}
    \caption{Portfolio value over time comparing ContestTrade against baseline strategies.}
    \label{fig:main_result}
\end{figure}

As shown in Table~\ref{tab:strategy_comparison} and Figure~\ref{fig:main_result}, ContestTrade achieves the highest performance among the evaluated methods in this backtest, with a CR of \textbf{52.80\%}, SR of \textbf{3.12}, and MDD of \textbf{12.41\%}. Compared with MASS, which employs fixed-agent cooperation without competitive selection, ContestTrade obtains higher profitability and risk-adjusted returns under the same trading constraints. ContestTrade also exceeds PPO and RSI\&KDJ in this setting, suggesting that dynamically filtering data factors and research agents can improve both return generation and risk control.

A closer examination across baseline categories illustrates where the gains arise. Traditional rule-based strategies (MACD, RSI\&KDJ) achieve modest returns (CR of 2.69\% and 8.19\%, respectively) with relatively controlled drawdowns, but their fixed rules cannot adapt to changing information quality. Machine learning approaches show mixed results: the standalone LGBM baseline loses money under our feature set, while LSTM achieves 8.34\% CR but with an elevated MDD of 29.56\%. Deep reinforcement learning methods (A2C, PPO) demonstrate stronger adaptability, with PPO reaching 15.07\% CR and the best baseline SR of 1.33. The multi-agent baseline MASS performs poorly in this A-share configuration, which is consistent with our motivation that unfiltered agent collaboration may amplify noisy signals. We emphasize that these are single-period backtest comparisons; without longer walk-forward evaluation and statistical tests, the results should be interpreted as evidence of promise rather than definitive dominance across market regimes.

To further investigate the source of performance, Table~\ref{tab:contest_effectiveness} presents the predictive power of each internal contest. The Data Analyst Contest achieved a mean Rank IC of 0.054 and ICIR of 0.13, while the Researcher Contest achieved Rank IC of 0.079 and ICIR of 0.18. These positive values indicate that contest scores contain useful ranking information for future factor and agent utility, but their magnitudes are modest and should be viewed as directional evidence. The results support the design choice of performance-based filtering, while motivating future work on stronger statistical validation and market-regime analysis.

\begin{table}[t]
\centering
\renewcommand{\arraystretch}{1.2}
\caption{Effectiveness of internal contest mechanisms.}
\label{tab:contest_effectiveness}
\begin{tabular}{l cc}
\toprule
\textbf{Component} & \textbf{Rank IC} & \textbf{ICIR} \\
\midrule
Data Analyst Contest & 0.054 & 0.13 \\
Researcher Contest & 0.079 & 0.18 \\
\bottomrule
\end{tabular}
\end{table}

\section{Ablation Studies}
\label{sec:ablation_studies}

To validate each component's effectiveness, we conduct ablation studies by systematically removing key mechanisms: (1) \textbf{w/o LLM Judge} disables LLM-based signal evaluation; (2) \textbf{w/o Contest - Researcher} removes the Research Team's competitive mechanism; (3) \textbf{w/o Contest - Data Analyst} disables the Data Team's factor selection contest; (4) \textbf{w/o Deep Research} removes the financial tool suite; and (5) \textbf{w/o All} removes all proposed mechanisms.

\begin{figure}[t]
    \centering
    \includegraphics[width=0.48\textwidth]{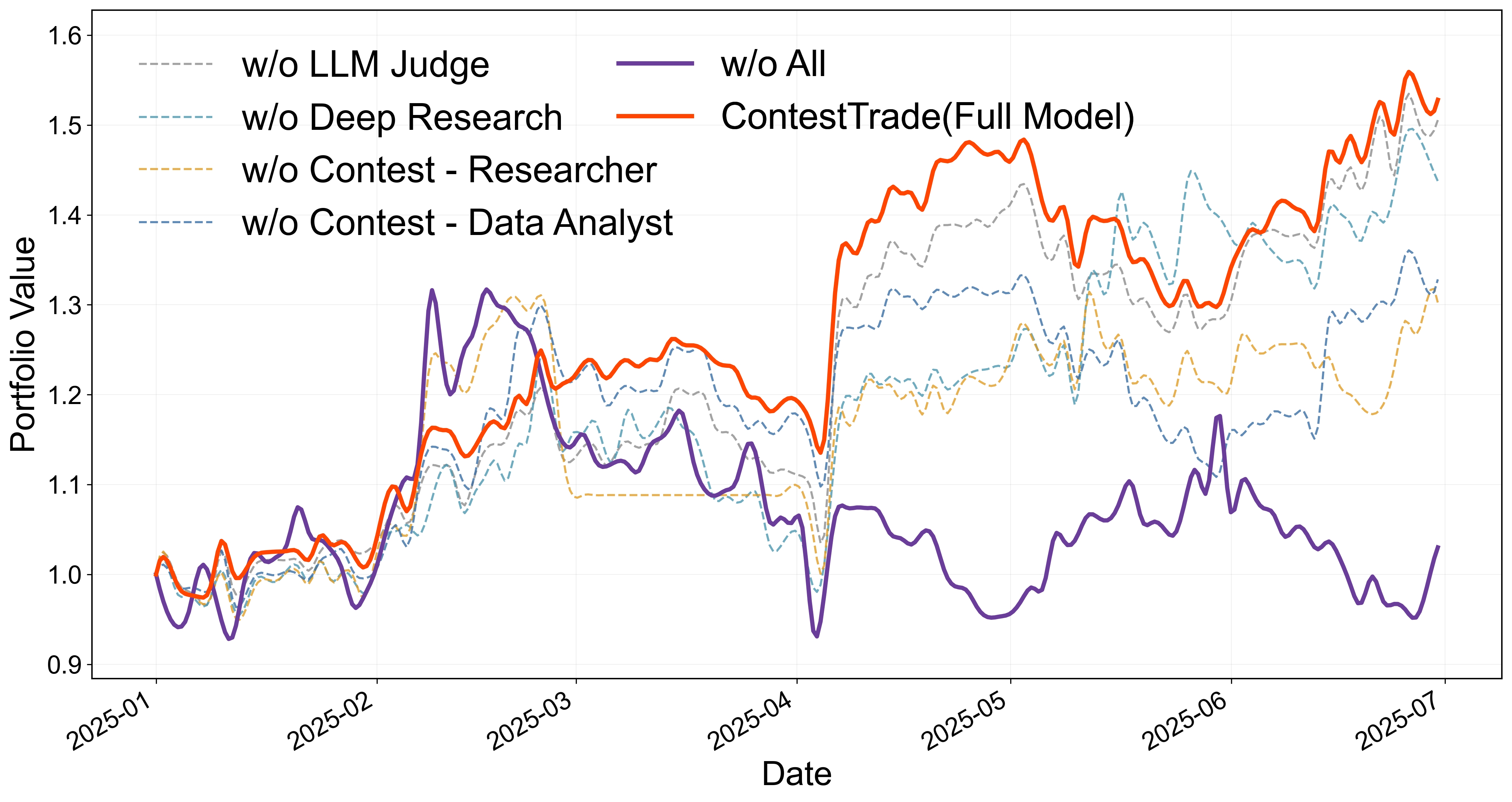}
    \caption{Ablation study: portfolio value over time for various configurations.}
    \label{fig:ablation_study_result}
\end{figure}

\begin{table}[t]
\centering
\renewcommand{\arraystretch}{1.2}
\caption{Ablation study results. ``w/o'' indicates component removal.}
\label{tab:ablation_study}
\begin{tabular}{l ccc}
\toprule
\textbf{Configuration} & \textbf{CR (\%)} & \textbf{SR} & \textbf{MDD (\%)} \\
\midrule
\textbf{ContestTrade (Full)} & \textbf{52.80} & \textbf{3.12} & \textbf{12.41} \\
\midrule
w/o LLM Judge & 50.55 & 2.57 & 13.48 \\
w/o Contest - Researcher & 32.83 & 1.78 & 16.70 \\
w/o Contest - Data Analyst & 42.85 & 2.01 & 13.47 \\
w/o Deep Research & 43.75 & 2.08 & 20.55 \\
w/o All & 3.01 & 0.07 & 26.63 \\
\bottomrule
\end{tabular}
\end{table}

As shown in Figure~\ref{fig:ablation_study_result} and Table~\ref{tab:ablation_study}, each component contributes distinctly in the evaluated period. Removing the LLM Judge causes SR to drop from 3.12 to 2.57 and MDD to increase to 13.48\%, suggesting that qualitative evidence checking helps risk-adjusted allocation even when contest rankings remain intact. The Researcher Contest is the most influential component in this ablation: its removal reduces CR from 52.80\% to 32.83\% and SR from 3.12 to 1.78, indicating that competitive evaluation among Research Agents is important for selecting investment theses. Similarly, removing the Data Analyst Contest degrades CR to 42.85\% and SR to 2.01, showing that factor-level quality control contributes beyond the research-stage contest.

The Deep Research mechanism shows a distinct impact profile: while its removal reduces CR to 43.75\%, it increases MDD to 20.55\%, the highest among single-component ablations. This pattern suggests that tool-augmented analysis may be particularly useful for identifying downside risks. Removing all contest and research mechanisms simultaneously leads to CR of 3.01\%, SR of 0.07, and MDD of 26.63\%, close to the passive CSI ALL Share benchmark. Overall, the ablation supports the complementarity of data filtering, researcher selection, LLM-based evidence judging, and tool use, although stronger conclusions require additional periods and repeated LLM sampling runs.

\section{Conclusion}
\label{sec:conclusion}

We introduced ContestTrade, a multi-agent framework for reducing inconsistent decision-making and market noise in LLM-based trading systems. Inspired by institutional investment practices, ContestTrade features specialized Data and Research teams with internal contest mechanisms that evaluate agents from delayed market feedback and allocate resources to outputs with positive predicted utility. In a controlled A-share backtest, ContestTrade achieves higher return and risk-adjusted performance than the evaluated baselines, and the positive Rank IC and ICIR values suggest that the contest scores contain useful predictive information.

Our contributions include three key innovations: (1) a dynamic multi-agent architecture with internal contest mechanisms that evaluate and rank agent outputs based on realized historical feedback; (2) a Deep Research methodology that equips agents with temporally constrained financial toolkits for multi-step investigative analysis; and (3) an information denoising pipeline through the Quantify-Predict-Allocate framework that transforms noisy market signals into actionable trading decisions under explicit trading rules.

We acknowledge several limitations of the current work. First, our experiments focus on the Chinese A-share market over a six-month test period; this is not sufficient to establish robustness across full market cycles, market regimes, or asset classes. Second, we report a single controlled backtest and do not yet provide confidence intervals from repeated LLM sampling, bootstrap tests, or longer walk-forward evaluations. Third, while we describe the temporal data protocol, real deployments would also need stricter audit trails for data-release timestamps, survivorship bias, market impact, liquidity, and capacity. Fourth, system performance depends on the underlying LLM and tool availability, and multi-agent tool-augmented research may be too costly for high-frequency settings. Finally, ContestTrade is a research prototype and should not be interpreted as investment advice.

Future work will address these limitations through longer walk-forward backtests, repeated-seed evaluation, transaction-cost and capacity sensitivity analysis, stronger cross-sectional quantitative baselines, and more detailed studies of LLM judge consistency. We also plan to evaluate international markets and adaptive contest windows that respond to detected regime shifts.

\section*{Acknowledgments}
This paper was supported by ``2025 Special Program for Supporting Innovative Development in Leading Industries (AI Track) under the High-Quality Industrial Development Initiative'' (Project Name: Finstep Finsmart Intelligent Service Platform; Project ID: 2025-GZL-RGZN-01024) from Shanghai Municipal Commission of Economy and Informatization, Shanghai, China.

\bibliographystyle{named}
\bibliography{references}

\end{document}